\documentstyle[sprocl,psfig]{article}

\input{psfig}
\def\be{\begin{equation}}
\def\ee{\end{equation}}
\def\bea{\begin{eqnarray}}
\def\eea{\end{eqnarray}}

\newcommand{\gfm}{ {\rm GeV/fm}^3}
\newcommand{\beq}{\begin{equation}}
\newcommand{\eeq}[1]{\label{#1} \end{equation}}

\newcommand{\lton}{\mathrel{\lower.9ex\hbox{$\stackrel{\displaystyle 
<}{\sim}$}}}
\newcommand{\een}{\end{enumerate}} \newcommand{\bit}{\begin{itemize}}
\newcommand{\eit}{\end{itemize}} \newcommand{\bc}{\begin{center}}
\newcommand{\ec}{\end{center}} 
\newcommand{\beqar}{\begin{eqnarray}}
\newcommand{\eeqar}[1]{\label{#1} \end{eqnarray}}

\newcommand{\vva}{{\bf v}_{\perp\alpha}}
\newcommand{\vxa}{{\bf x}_{\perp\alpha}}
\newcommand{\vxp}{{\bf x}_\perp}
\newcommand{\vpa}{{\bf p}_{\perp\alpha}}

\newcommand{\pa}{p_{\perp\alpha}}

\begin{document}
\begin{flushright}
CU-TP-757\end{flushright}
\title{TRANSVERSE SHOCKS IN THE TURBULENT GLUON PLASMA PRODUCED
IN ULTRA-RELATIVISTIC A+A}

\author{ M. GYULASSY \footnote{Speaker at Int. Conf. Nucl. Phys.,Wilderness/
George, SA, March 10-16, 1996}, D. H.\ RISCHKE,
and B. ZHANG}

\address{Physics Department, Pupin
Physics Laboratories, Columbia University,\\ 
538 W 120th Street, New York, NY 10027, U.S.A.}

\maketitle\abstracts{
Mini-jet production in ultra-relativistic nuclear
collisions leads to initial conditions characterized by 
large fluctuations of the local energy density (hot spots) and 
of the collective flow field (turbulence). 
Assuming that local equilibrium is reached on a small time scale,
$\sim 0.5$ fm/c, the transverse evolution of those initial conditions
is computed using hydrodynamics. We find that a new class
of collective flow phenomena (hadronic volcanoes) could
arise under such conditions. This could be
observable via enhanced azimuthal fluctuations 
of the transverse energy flow, $d^2E_\perp/d\phi dy$.
}

\section{Introduction}
At energies $\sqrt{s} > 100$ AGeV, mini-jet production
in central nuclear collisions is expected \cite{eskola}
to be the primary dynamical source of 
a plasma of gluons with an initial energy density
an order of magnitude above the deconfinement
and chiral symmetry scale, $\epsilon_c \sim 1\, \gfm$.
Many observable consequences of the formation
of this new phase of matter have been 
predicted based on a variety of assumptions \cite{qm93},
and experiments are currently under construction
to search for evidence for that so called quark-gluon plasma (QGP)
at the Relativistic Heavy--Ion Collider (RHIC) at Brookhaven.
Evidently, signatures depend sensitively on the assumed 
ensemble of initial conditions generated in such collisions.
Most often it is assumed for simplicity
that a homogeneous, cylindrically symmetric,
and longitudinally boost-invariant quark-gluon plasma is created 
and thus that signatures
can be computed ignoring fluctuations of the initial conditions
themselves. 

In this talk we point out, however, that the mini-jet formation mechanism
leads to a rather inhomogeneous and turbulent
ensemble of initial conditions that is characterized by 
a wide fluctuation spectrum of the local energy density (hot spots) and 
of the collective flow field (turbulence).
In this case, some of the proposed signatures will be
washed out while new ones will certainly arise.
We show below that in fact a new type of collective ``shock'' phenomena
may occur under these conditions 
that could be readily observed as unusual transverse
energy fluctuations.

This topic is well suited to this symposium honoring
the 60th birthday of Professor Greiner because he was
the first to propose in 1974 the ``shocking'' idea
that nuclear shock waves may be produced in nuclear collisions
and serve as an ideal probe of the 
equation of state of dense matter \cite{shock}. 
Since the experimental discovery of nuclear shocks in the 1 AGeV range
in 1984 and in the AGS energy range last year \footnote{see
talks by J.\ Symons, J.\ Stachel, and P.\ Braun-Munzinger in 
these proceedings.}, much theoretical and experimental effort
has been devoted to this topic
over a large  energy range:
$E_{lab}=100$ AMeV to 100 ATeV. The present work focuses on 
a new class of shock phenomena that could occur in quark-gluon plasmas
near the top end of that energy scale.

To illustrate the novel type of collective flow patterns we
have in mind, the hydrodynamic \cite{dirk} evolution
of initial conditions consisting of three static cylindrical ``hot spots''
are shown in Fig.\ 1. 
\begin{figure}[h]
\hspace{1.5in}
\psfig{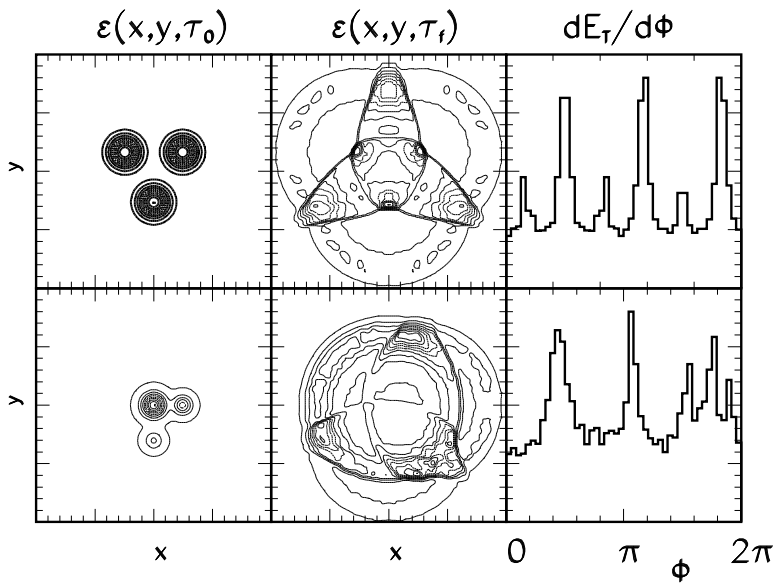}
\caption{Hadronic volcanoes erupt due to shock
formation in regions where expanding shells of matter 
evolved from inhomogeneous
``hot spots'' initial conditions (left panels) intersect.
The signature of such volcanoes is the strong
anisotropic azimuthal angle dependence
of the transverse energy flow
$dE_T/d\phi$ (right panels).
Contours of constant energy density,
$\epsilon(x,y,\tau)$, at the breakup time $\tau_f$
are shown in the middle panels.
The top row shows the evolution from a symmetric initial configuration
of hot spots. The bottom rows shows the evolution from a more 
realistic asymmetric configuration.
}
\end{figure}

An ideal gas equation of state
($p=\epsilon/3$) is assumed.
Shock waves are seen to be created where the three  rapidly expanding shells
of matter intersect at six points.  
The collective acceleration of matter in those shock
zones can be clearly seen 
as  strong peaks in the transverse energy flow.
For the symmetric configuration (Fig.\ 1a), a ``mercedes'' pattern emerges 
(Fig.\ 1b), leading to three primary and three secondary shocks in Fig.\ 1c.
For the asymmetric case in Figs.\ 1d,e
the azimuthal angle dependence of the  shock peaks is spread out
and modified.  The search for analogous 
enhanced anisotropic  transverse energy fluctuations (or hadronic 
volcanoes \footnote{T.D. Lee first used this term for anisotropy
that could arise due to surface instabilities. In our case
the ``volcanoes'' are dynamically generated from
specific inhomogeneous initial conditions.}) as 
a function of rapidity and azimuth is the subject of the 
following discussion.

\section{HIJING Initial Conditions}

Mini-jet production is thought to be the dominant
mechanism controlling the initial conditions in $A+A$
because the pQCD inclusive cross section
for jets with moderate $p_\perp > p_0\sim 1-2$ GeV rises to the 
value $\sigma_{jet}(p_0) > 10$ mb at RHIC energies.
Compelling evidence for pQCD mini-jet dynamics has been
observed in $pp$ and $p\bar{p}$
reactions at collider energies \cite{wang}.

Due to the nuclear geometry, 
the total number of mini-jet gluons produced in central $A+A$ collisions
is expected to be $\sim A^{4/3}\sim 10^3$ times larger than in $pp$
collisions. This simple geometric enhancement causes the rapidity density
of mini-jet gluons to reach $dN_g/dy\sim 300-600$ in $Au+Au$
depending on the mini-jet transverse momentum cut-off scale
$p_0=1-2$ GeV/c. The hadronization of this mini-jet gluon plasma via the
string fragmentation mechanism in HIJING approximately doubles 
the final hadron transverse energy distribution due to the pedestal or
``string'' effect. In the pre-hadronization phase of the evolution
we therefore include a ``soft'' component in terms of a background
gas of soft gluons normalized such that
together with the mini-jet component the final transverse energy distribution
predicted by HIJING is reproduced.

The average initial energy density of the mini-jet gluon
plasma can be estimated using the Bjorken formula:
$\langle {\epsilon}(\tau)\rangle\approx (p_0/\tau_0 \pi
R^2)\; dN_g/dy \;(\tau_0/\tau)\sim 40\; (\tau_0/\tau)\; \gfm $. 
This applies until the thermalization time is reached and work due
to expansion must be considered.

To see that the initial conditions are in fact highly inhomogeneous
and are dominated by a few ``hot spots'',
we must calculate the {\em local} energy density
$\epsilon(\tau,\vxp)$ instead of averaging over the transverse coordinate
as in the Bjorken formula. The local energy density and transverse 
momentum density at $z=0$ (the $y_{cm}=0$ frame)
can be computed from the HIJING list of produced gluons as
\beqar
\left(\begin{array}{c}
\epsilon(\tau,\vxp) \\
\vec{m}(\tau,\vxp)
\end{array}\right)
= \sum_{\alpha}\left(\begin{array}{c}
1\\ \vva
\end{array}\right) 
\frac{\tau 
\pa^3}{1+(\tau \pa)^2}~\delta^2(\vxp-\vxa(\tau))~
\delta(y_\alpha)
\; \; .\eeqar{xebj}
The sum is over the produced gluons with 
transverse and longitudinal momentum $(\vpa, p_{z\alpha}=
p_{\perp\alpha}\sinh y_\alpha)$.
The longitudinal and transverse coordinates of the production, 
$(z_\alpha=0,\vxa$), are determined from the coordinates of the 
binary nucleon-nucleon collision from which the gluons
originate. The above formula takes into account the free streaming of gluons
not only along the $z$ direction via the volume element $\tau \Delta y$
but also in the transverse direction via $\vxa(t)=\vxa+ \vva\tau $,
where $\vva=\vpa/p_\alpha $.
The factor $1/(1+(\hbar/\tau \pa)^2)$
is the formation probability \cite{GyuWa} of the gluon in the comoving frame.
High--$p_T$ gluons are produced first and gluons with
lower $p_T$ later according to the uncertainty principle
leading to the so-called inside-outside picture of the dynamics.
 Before $\tau\sim \hbar/\pa$
that component of the radiation field is still part of the coherent
Weizs\"acker-Williams field of the passing nuclei \cite{mclerran}.

The expression (\ref{xebj}), when averaged over transverse coordinates,
recovers the Bjorken expression 
$\langle \epsilon(\tau)\rangle\approx \epsilon_0 (\tau_0/\tau)$
with  vanishing average transverse momentum density, $\langle
\vec{m}(\tau)\rangle=0$.

To study the transverse coordinate dependence of the
initial conditions, we must specify a transverse, $\Delta r_\perp$,
and longitudinal, $\Delta y$, resolution scale. 
The densities coarse-grained on that resolution scale are 
obtained from (\ref{xebj}) substituting
\beq
\delta^2(\vxp-\vxa(\tau))\, \delta(y_\alpha) \rightarrow
\frac{\Theta(\Delta y/2 - |y_{\alpha}|)
}{\Delta r_\perp^2 \Delta y}\,
\Theta(\Delta r_\perp/2- |\vxa(\tau)-\vxp|)
\; \; .\eeq{smear}
To determine the relevant resolution scale,
we note that at time $\tau$ the local horizon for any gluon
in the comoving $(y_g=0)$ frame has a radius $c\tau$.
Thus at the thermalization time, $\tau_{th}$,
each gluon can be influenced by only a small neighborhood of radius
$c\tau_{th}\approx 0.5$ fm  of the mini-jet plasma.
We take the transverse resolution scale to be the maximal
causally connected diameter, $\Delta r_\perp=2c\tau_{th}\approx 1$ fm
and the rapidity width to be $\Delta y=1$. Gluons moving with
larger relative rapidity $y$ are produced later, $\tau(y,p_\perp)
\sim \cosh(y)/p_\perp$  due to time dilation
at the boundary of the local horizon.
The above choice of the resolution scale is 
the most optimistic from the point of view of {\em minimizing} fluctuations
of the thermodynamic variables between
the causally disconnected domains in the transverse plane.
We note that the number, $N_d(\tau)=(R/\tau)^2$, 
of such causally disconnected domains
in a  nuclear area, $\pi R^2$, is initially very large.
Even at later times, $\tau \sim 3$ fm/c,
when the mean energy density falls below 
$\epsilon_c$, several disconnected domains remain.

The soft component is modeled by $\approx 1700$ low--$p_\perp$
gluons per unit rapidity
with a Gaussian transverse momentum distribution with
rms $p_\perp=0.3$ GeV/c. This leads to  about 500 GeV of transverse energy per
unit rapidity as  needed to reproduce the HIJING final $dE_T/dy$.
However, these soft gluons
only add a relatively small, approximately homogeneous contribution
to the energy density, $\epsilon_s\approx 4\, \gfm$, at $\tau_{th}=0.5$ fm/c.

\section{Hot Spots and Turbulence}

\begin{figure}[h]
\vspace{-0.5cm}
\hspace{0.5in}
\psfig{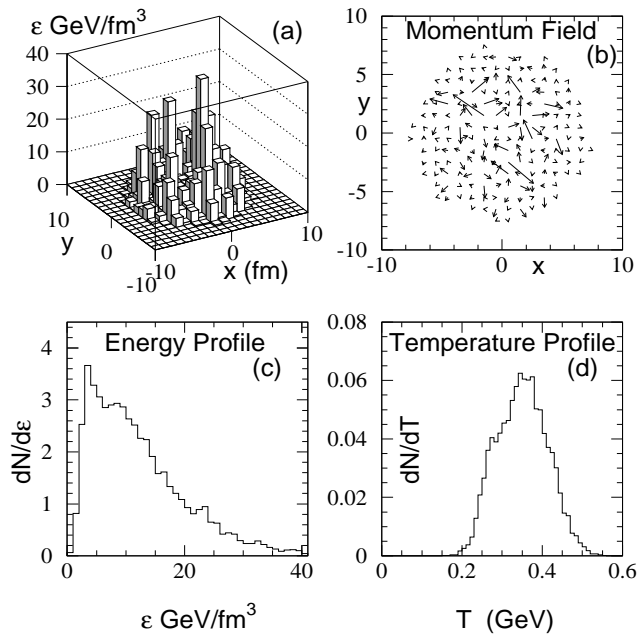}
\caption{(a) Energy density distribution of hot spots
in a typical  HIJING event. Note the large fluctuations
of the initial energy density across the transverse plane.
(b) Momentum density fluctuations indicate considerable initial state
turbulence. (c) Distribution of energy density on a
$\Delta r_\perp=1$ fm 
transverse coordinate resolution scale averaged over 200 HIJING events.
Note that this resolution scale corresponds to the maximal allowed one
by causality at proper time, $t=0.5$ fm/c.
(d) The thermal profile distribution on the same resolution scale
has a rather large width. The mean and shape of course depend
also on the equation of state (assumed here to be an ideal gas).
}
\end{figure}

The inhomogeneous nature of the mini-jet plasma
initial conditions in central $Au+Au$ collisions at RHIC
is illustrated in Fig.\ 2. In Fig.\ 2a the energy density profile
at an assumed thermalization time
$\tau=0.5$ fm/c  of a typical HIJING ($b=0$) event
at $\sqrt{s}=200$ AGeV is shown. The plasma in the central rapidity slice
($\Delta y_{cm}=1$) exhibits
large fluctuations (hot spots) because the average number of
hard mini-jets produced per nucleon is only $\sim 1$ per unit rapidity
at RHIC energies. The hot spots are also associated with
a chaotic velocity field (turbulence) 
as seen in Fig.\ 2b.
When averaged over many events the distribution
of the proper energy density and temperature on a transverse
resolution scale $\Delta r_\perp=1$ fm are shown in Figs.\ 2c,d.
Note that at that formation time the plasma cannot be characterized
by a unique temperature. In fact the widths of those distributions
 which are controlled by the
Glauber geometry of finite nuclei and the size of the mini-jet
cross section obviously cannot be neglected.
Signatures
of plasma formation will differ considerably
in this turbulent gluon scenario than in the
conventional homogeneous hot-glue scenario.
For example high--$p_T$ direct photon production and 
heavy--quark production
will be greatly enhanced in the hot spots \cite{hotspot}.
In the next section we find that hadronic volcanoes similar
to those in Fig.\ 1 also arise.

\section{Hadronic Volcanoes and Transverse Energy Fluctuations}

In Figure 3 the azimuthal angle dependence of the
transverse energy flow, $dE_\perp/ dyd\phi$, is shown for a typical HIJING
initial condition. The left panels (Figs.\ 3a,c)
correspond to an ideal gas equation of state, while the right panels (Figs.\
3b,d) correspond to a first order Bag model equation of state \cite{dirk}.
The top two panels evolve the initial distribution in Fig.\ 2a
with the momentum field in Fig.\ 2b 
artifically set to zero ($\vec{m}(\vxp,\tau_0)=0$). 
This static inhomogeneous initial condition is most similar to the one in
Fig.\ 1. The initial isotropic nature of the thermally folded
transverse energy distribution is seen by the dashed lines in Figs.\ 3a,b.
The lower panels (Figs.\ 3c,d) correspond to the evolution of
the turbulent initial condition in Figs.\ 2a,b. The turbulence of the
initial condition is revealed by the anisotropy of the dashed curves.

In all cases the magnitude of the transverse energy decreases
with time due to the work done by the plasma
undergoing longitudinal (Bjorken) expansion.
More work is done by an ideal gas case than a plasma
with first order transition because
the latter equation of state is softer \cite{dirk}.
In the static hot spot case the production of hadronic volcanoes is
indeed similar to Fig.\ 1. In the turbulent case,
the effect is obscured somewhat by the large initial anisotropy.
Nevertheless, in the turbulent case also
several sharper peaks appear.
Evidence for the collective origin of the hadronic volcanoes
in the turbulent  case
must be looked for in higher order $E_T$ correlation analysis,
e.g., $\langle E_T(\phi) E_T(0) \rangle$. 
For a more detailed discussion of these results we refer
to Ref.\ \cite{hotspot}.

\begin{figure}[h]
\hspace{0.5in}
\psfig{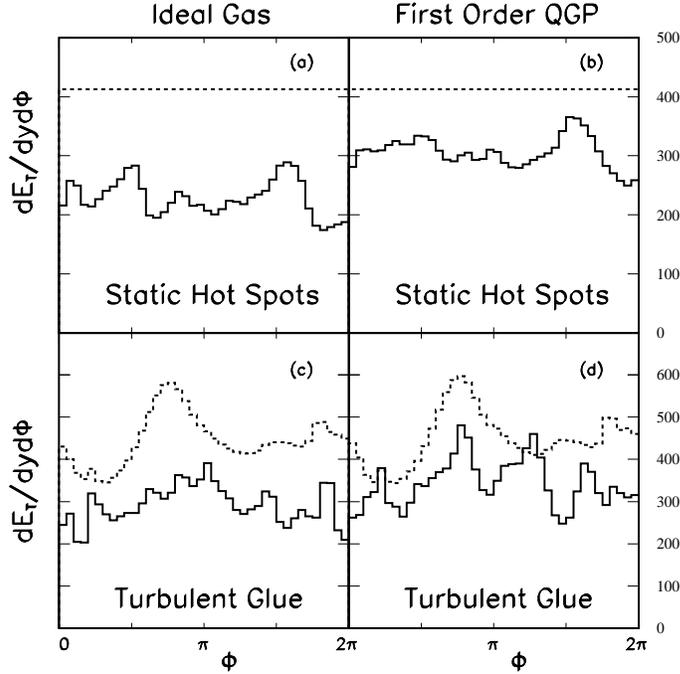}
\caption{The evolution of
transverse energy $dE_T/dyd\phi$ at $y=0$ is shown for different
inhomogeneous HIJING initial conditions and equations of state.
In parts (a,b) the fluid velocity field shown in Fig.\ 2b
is set to zero while in parts (c,d) the
turbulent velocity field is included. 
The evolution in parts (a,c) assumes an ideal
gas law $p=\epsilon/3$, while in parts
(b,d) a  first order Bag model transition is included.
The dashed curve is the initial transverse energy distribution
at the onset of hydrodynamic evolution. In the static hot spot
cases the initial transverse energy is uniformly distributed in azimuth,
while in the turbulent case large initial fluctuations are generated by
the mini-jets.
The solid curves show the final transverse energy distribution.
}

\end{figure}

\section*{Acknowledgments}
This work was supported by the Director, Office of Energy
Research, Division of Nuclear Physics of the Office of High Energy and Nuclear
Physics of the U.S.\ Department of Energy under Contract No.\ 
DE-FG-02-93ER-40764. D.H.R.\ was partially supported by the 
Alexander von Humboldt--Stiftung under 
the Feodor--Lynen program.

\end{document}